\documentclass{ws-ijmpd}
\newcommand{\et}{{\it et al.}}
\newcommand{\apj}{{\it ApJ}}
\newcommand{\mn}{{\it MNRAS}}
\newcommand{\aap}{{\it A\&A}}
\begin{document}

\markboth{Arieh K\"onigl}
{Relativistic Jets: Open Problems and Challenges}

%
\catchline{}{}{}{}{}
%

\title{RELATIVISTIC JETS: OPEN PROBLEMS AND CHALLENGES}

\author{ARIEH K\"ONIGL}

\address{Department of Astronomy \& Astrophysics, The University of
Chicago\\
5640 S Ellis Ave, Chicago, IL 60637, USA\\
akonigl@uchicago.edu}

\maketitle

\begin{history}
\received{Day Month Year}
\revised{Day Month Year}
\comby{Managing Editor}
\end{history}

\begin{abstract}
Outstanding questions in the study of relativistic jets in their various
astrophysical settings are discussed in the context of a general
dynamical model.
\end{abstract}


\section{Introduction}
\label{sec:intro}

There are a number of possible approaches to how the key open questions
in the study of high-energy phenomena in relativistic outflows can be
articulated. The one adopted here seizes on the basic dynamical model
that is commonly believed to account for the acceleration and
collimation of relativistic jets as a useful framework for this
discussion. I first present the main elements of this model in the
context of the different astrophysical settings in which relativistic
jets occur. I then apply this model to the interpretation of the
observational data and consider how it might be further tested and
refined. Finally, I review how the connection between accretion and
outflows -- a fundamental property of cosmic jets -- is manifested in
these sources.

The ``canonical'' relativistic jet sources encompass gamma-ray bursts
(GRBs), active galactic nuclei (AGNs), and X-ray binaries (XRBs). The
bulk Lorentz factors implicated in GRB outflows are deduced from
arguments involving opacity to pair creation or electron scattering for
the high-energy photons\cite{LS01} and typically lie in the range
$\Gamma \sim 10^2-10^3$ for long/soft bursts. A prominent recent example
is GRB 080916C, for which $\Gamma > 600$ has been inferred in this
way.\cite{Abd09} In the pre-{\it Swift} era, several well-studied
sources appeared to exhibit panchromatic breaks in their afterglow
light curves on time scales (at the source) of $\gtrsim 1\,$day, from
which characteristic jet opening half-angles $\theta_{\rm j} \sim
2^{\circ}-5^{\circ}$ have been deduced.\cite{ZKK06} However, {\it
Swift}-era bursts have exhibited a more complex behavior that has cast
some doubt on this interpretation.\cite{Rac09} In fact, in some cases
jet breaks were argued to have occurred on much shorter
time scales,\cite{Kam09}\cdash\cite{Pan07} implying $\theta_{\rm j}
\lesssim 1^{\circ}$. GRB outflows are likely powered by the extraction
of rotational energy from a newly formed stellar-mass black hole or a
rapidly rotating neutron star, or from a surrounding debris disk.

In the case of AGNs, jets are actually imaged in radio through X-ray
wavelengths, and their relativistic motions in the blazar class of
objects are inferred directly from apparent superluminal motions of
radio emission features on scales $\lesssim 1\,$pc. Typical apparent
speeds are $v_{\rm apparent}/c \sim 10$ although values as high as $\sim
50$ have been measured, and it is likely that the bulk Lorentz factors
are of the same order.\cite{Lis09} These inferences are consistent with
the values ($\Gamma \gtrsim 50$) deduced from pair-opacity arguments in
the BL Lac objects PKS 2155-304 and Mrk 501 on the basis of the rapid
($3-5\,$min) TeV-flux variability that they exhibited,\cite{BFR08}
although in these particular sources the radio emission knots have much
lower apparent speeds and the implied bulk Lorentz factors may not be
representative of the jet as a whole.\cite{Ghi09b}\cdash\cite{GUB09}
Although AGN jets evidently also contain a nonrelativistic outflow
component, there are indications that the relativistic component
persists to large (kpc to Mpc) scales. These include the measurement of
apparent superluminal motions at a projected distance of at least
$150\,$pc in the 3C 120 jet,\cite{Wal01} the results of spectral
modeling of the lobes of powerful radio sources, which suggest that the
flow might be decelerating from relativistic speeds in the jet
termination shock,\cite{GK03} and the detection of extended X-ray
emission in quasar jets (if interpreted as beamed, inverse-Compton
scattered, cosmic-microwave-background radiation; see
Ref.~\refcite{Wor09}).

Apparent superluminal radio-component motions are also detected in X-ray
emitting neutron-star and black-hole binaries, and although the typical
speeds ($\sim 2-5\, c$) appear to be lower than those of AGNs, a value
$\ge 12\, c$ has already been measured in Cir X-1 (a neutron-star
binary; Ref.~\refcite{Tud08}). In all these astrophysical realizations
of a collimated relativistic outflow it is believed that a {\it magnetic
field}\/ provides the most plausible means of extracting rotational
energy from the source (a black hole, neutron star, or accretion disk).
A magnetic field (especially if it assumes a large-scale, ordered form)
can also naturally guide, collimate, and accelerate these
flows.\cite{Bla02} However, thermal energy may contribute to the
jet acceleration in GRB sources (through neutrino emission from the
disk) and (in the case of long/soft bursts that originate in collapsed
massive stars) to the re-acceleration of shocked jet material when it
emerges from the progenitor star.

\section{Magnetic Acceleration and Collimation of Relativistic Outflows}
\label{sec:model}

The dynamical model adopted in the ensuing discussion is based on exact
(axisymmetric) solutions to the equations of special-relativistic, ideal
magnetohydrodynamics (MHD) obtained in semi-analytic form (assuming time
independence and employing radial self-similarity) and through
multi-scale numerical simulations that attain a quasi-steady
state.\cite{LCB92}\cdash\cite{Lyub09} For definiteness, it is assumed
that the outflow originates in an accretion disk, but a flow that taps
instead the rotational energy of the central compact object would be
qualitatively similar. As the bulk of the jet acceleration generally
occurs at large distances from the origin (see
Sec.~\ref{subsec:characteristics}), the effects of the central source's
gravity can be neglected. However, the details of the flow behavior in
the immediate vicinity of the compact object may require a fully
general-relativistic treatment.\cite{Lyut09} The disk is assumed to be
threaded by a large-scale, ordered magnetic field with an ``even''
symmetry (i.e. having only a $z$ component at the midplane, using
cylindrical coordinates $r,\; \phi,\; z$), although in principle a
small-scale, tangled field could also accelerate and collimate the
flow.\cite{HB00}\cdash\cite{Li02} The field transfers angular momentum
and energy from the accretion flow to the jet. The outflow is envisioned
to be Poynting flux-dominated at its base, with the acceleration
effected through the gradual transformation of magnetic energy into
kinetic energy as the jet propagates away from the origin.

\subsection{General characteristics}
\label{subsec:characteristics}

For the steady-state problem one jointly solves the Bernoulli and
transfield (Grad-Shafranov) equations, representing, respectively, the
force balance along and across poloidal streamlines.  The detailed
properties of the derived solutions depend on the particulars of the
boundary conditions (for example, the distributions of magnetic flux and
angular velocity at the source). However, when one follows the solutions
to sufficiently large scales (the ``asymptotic'' regime) one can
identify several generic properties of such flows. In particular, it is
found that the magnetic acceleration mechanism is quite efficient,
typically leading to a rough equipartition between the Poynting and
kinetic-energy fluxes. The bulk of the acceleration is due to the
gradient along the flow of the magnetic pressure associated with the
azimuthal field component ($B_\phi$). MHD acceleration is in general
spatially extended, with the bulk of the increase of $\Gamma$ occurring
over distances that are much larger than the characteristic scale of a
purely hydrodynamical (i.e. thermal) acceleration (which is of the order
of the size of the region where the accelerated gas is initially
confined).

The magnetic acceleration process is intimately tied to the shape of the
poloidal field lines, which needs to be derived (using the
Grad-Shafranov equation) simultaneously with the kinematic properties of
the flow (obtained from the Bernoulli equation). In particular, in order
for the flow to accelerate, the transverse distance between neighboring
field lines must increase faster than their cylindrical radius.  This
can occur naturally in current-carrying jets, which develop a
cylindrical core (due to compression by the $B_\phi$ hoop stress) even
as the outer parts of the flow collimate more slowly. The fact that
magnetic acceleration and collimation go hand in hand is expressed by
the following general asymptotic result: $\Gamma\tan{\theta_{\rm v}} =
1/\sqrt{b-1}$, where $\theta_{\rm v}\equiv \arctan(dr/dz)$ is the local
opening half-angle of the magnetic flux surface and the parameter $b$
describes the local field-line shape ($z\propto r^b$, with $1<b\le 2$).
This implies that $\Gamma\tan{\theta_{\rm v}} \approx \Gamma\theta_{\rm
v}$ is usually $\simeq 1$ for ``collimation accelerated''
jets.\cite{Kom09}\cdash\cite{Lyub09}

The asymptotic Lorentz factor $\Gamma(\theta)$ of the model jets
typically peaks away from the axis. This is because the acceleration is
due mainly to the gradient along the flow of the magnetic pressure
associated with $B_\phi$, and $B_\phi$ vanishes on the axis by
symmetry. However, the kinetic power per unit solid angle
$\epsilon(\theta)$ usually does peak at $\theta \approx 0$, reflecting
the higher density attained in the collimated core. These generic
distributions are quite distinct from those commonly adopted in
phenomenological jet models --- for example, the ``universal'' model of
GRB outflows ($\epsilon \propto \theta^{-2}$) and the ``hollow cone''
model (in which both $\Gamma(\theta)$ {\it and}\/
$\epsilon(\theta)$ peak away from the axis).

\subsection{Confinement and collimation}
\label{subsec:confine}

The collimation by the magnetic hoop stress of the innermost streamlines
of a current-carrying jet relative to the outer streamlines represents
the {\it self-collimation}\/ property of magnetized outflows. However, a
magnetized jet {\it cannot}\/ be ``self-confined'': confinement requires
an ambient (thermal, magnetic, or ram) pressure.

The spatial distribution of the confining external pressure (e.g.
$p_{\rm ext}(z) \propto z^{-{\alpha}}$) determines
the shape of the flow boundary (e.g. $z\propto r^{\beta}$) and hence the
acceleration efficiency (see Sec.~\ref{subsec:characteristics}). The
correspondence between the exponents $\alpha$ and $\beta$ can be
summarized as follows (see
Refs.~\refcite{Tch08}, \refcite{Kom09} and~\refcite{Lyub09}): 
\begin{romanlist}[(ii)]
\item 
${\alpha} < 2\ {\Leftrightarrow}\ {\beta}=4/\alpha >2$
\item 
${\alpha} = 2\  {\Leftrightarrow}\ 1<{\beta}\le2$
\item 
${\alpha} > 2\ {\Leftrightarrow}\ {\beta} = 1$ (asymptotically)\, ,
\end{romanlist}
where the specific value of $\beta$ for $\alpha=2$ is determined by the
normalization of $p_{\rm ext}$ (i.e. by its magnitude at the base
of the flow and not just by the value of its power-law exponent). These
relationships demonstrate that efficiently accelerating jets are
characterized by paraboloidal streamlines.

In order for the bulk of the flow to be efficiently accelerated there
has to be {\it causal connectivity}\/ across the flow: the inner and
outer streamlines must be able to communicate (through the propagation
of fast-magnetosonic waves) on the radial expansion time.  Causal
connectivity {\it cannot}\/ be maintained if $\alpha > 2$: in that case
the boundary asymptotically approaches a conical shape --- corresponding
to ``free'' (ballistic) expansion --- and significant acceleration
occurs only near the axis. Notwithstanding this conclusion, a magnetized
relativistic jet can undergo significant acceleration if it loses
pressure support (i.e. $\alpha$ increases from a value $\le 2$ to a
value $> 2$) when it is already super--fast-magnetosonic.\cite{Tch10}
This acceleration is effected by a rarefaction wave that communicates
the loss of pressure support at the jet boundary to the gas in its
interior,\cite{KVK10} although the ``rarefaction acceleration'' process
can also be described in terms of the transverse distance between
neighboring field lines increasing faster than their cylindrical radius
--- just like the ``collimation acceleration'' mechanism discussed in
Sec.~\ref{subsec:characteristics}. Because of the way that velocities
add in special relativity, the transverse acceleration induced by a
rarefaction wave that propagates into an already relativistic flow can
lead to a significant boost in the jet Lorentz factor. Although this
mechanism is applicable also in the purely hydrodynamical case, it is
particularly effective when the flow is
magnetized.\cite{AR06}\cdash\cite{Miz08}

External confinement is especially important for relativistic flows
since, as $\Gamma$ becomes $\gg 1$, the increased inertia and the growth
of the electric force (which nearly cancels the transverse magnetic
force) reduce the collimation efficiency. One possible confining agent is a
{\it disk wind}, which could itself be driven hydromagnetically. This
leads naturally to a {\it two-component} jet model, with a fast, narrow
``spine'' surrounded by a wider, slower, outflow. The fast component
could be associated with the central object -- for example, a
Blandford-Znajek jet from the black-hole ergosphere\cite{LE00} --- or
else it could correspond to the innermost disk outflow. In the case of
GRBs, the disk outflow could itself be relativistic ($\Gamma \gtrsim
10$) and might account for the bulk of the afterglow
emission.\cite{PKG05}\cdash\cite{Rac08} In the case of AGNs, this type
of model has long been advocated\cite{SPA89}\cdash\cite{SO02} given that
distinct relativistic and nonrelativistic outflow components are
directly observed. The relative contribution of two such components --
attributed to the black hole and disk, respectively -- is one possible
explanation of the FR I/FR II dichotomy of extragalactic radio
sources.\cite{SSL07} A potential consistency check of this hypothesis
might be a search for broad iron lines in the X-ray spectra of FR II
galaxies, from which a high rotation parameter for the central black
hole might be inferred.\cite{Fab00} The practical implementation of such
a test would, however, be challenging.\cite{Bal07}\cdash\cite{Lar08}

\section{Confrontation of the Model with Observations}
\label{sec:confront}

\subsection{GRB sources}
\label{subsec:GRB}

Ideal-MHD numerical simulations of magnetically accelerated
jets\cite{Kom09} indicate that GRB outflows can attain Lorentz factors
$\Gamma \gtrsim 10^2$ on scales $\sim 10^{10}-10^{12}\,$cm, consistent
with the possibility that long/soft GRB jets are accelerated within the
envelopes of collapsing massive stars (which provide the requisite
confinement). It is furthermore found that $\Gamma \gtrsim 30$ (a
fiducial lower limit on the Lorentz factor in short/hard GRB jets;
Ref.~\refcite{Nak07}) can be attained on scales $\sim 9\times
10^{8}-3\times 10^{10}\,$cm, consistent with the possibility that in
this case the confinement is provided instead by a moderately
relativistic wind from a central accretion disk that forms during a
binary (two neutron stars or neutron-star/black-hole) merger. If the
initial magnetizations of short/hard and long/soft GRB outflows are
comparable, this scenario provides a plausible explanation of the
finding (from the best available current data) that short/hard GRB jets
are on average less relativistic than the ones associated with long/soft
GRBs, and it leads to the prediction that short/hard GRB outflows should
also be less well collimated, on average, than their long/soft GRB
counterparts.

The relation $\Gamma \theta_{\rm v} \approx 1$ obtained for the
``collimation acceleration'' mechanism implies $\theta_{\rm v} \lesssim
1^\circ$ for $\Gamma \gtrsim 100$. As noted in Sec.~\ref{sec:intro},
this is consistent with the inferences from some early-afterglow
observations but it does not fit the pre-{\it Swift}\/ standard model,
in which $\Gamma \theta_{\rm v} \gg 1$ is deduced from apparent
panchromatic breaks in the afterglow light curve at later times. In the
case of long/soft GRBs, the latter behavior can be attributed to the
``rarefaction acceleration'' mechanism, which is likely to operate when
the outflow becomes unconfined upon emerging from the envelope of the
progenitor star.\cite{Tch10}\cdash\cite{KVK10} Specifically, if the flow
arrives at the stellar surface with $\Gamma \gtrsim 10^2$ and $\Gamma
\theta_{\rm v} \approx 1$, the ``rarefaction acceleration'' mechanism
could increase $\Gamma$ by a factor of a few (with $\theta_{\rm v}$ also
increasing, but by no more than $\sim 1/\Gamma$).  It is unclear how
this scenario could accommodate very early jet breaks, although further
observations are needed to verify that such breaks really exist. Given
that short/hard GRB outflows evidently do not propagate through a
stellar envelope, the above mechanism is probably not relevant to these
sources. This picture therefore leads to the prediction that there
should be few (if any) ``latish'' jet breaks in the afterglows of
short/hard GRBs, which could be tested as more such afterglows are
observed. Yet another consistency check arises from the findings that
rarefaction acceleration is most effective in highly magnetized jets and
that the latter are unlikely to form a hot cocoon of shocked gas as they
advance toward the stellar surface:\cite{KB07} the above picture may
thus be incompatible with models\cite{TWM09} that attribute some of the
emission properties of a source like GRB 080916C to the presence of such
a cocoon. In contrast with the magnetic acceleration scenario, late-time
jet breaks are not an issue for the fireball model, in which there is no
inherent upper bound on $\Gamma \theta_{\rm v}$.

\subsection{AGN jets}
\label{subsec:AGN}

A growing body of data indicates that relativistic AGN jets undergo the
bulk of their acceleration on spatial scales not much smaller than $\sim
0.1\,$pc, which are much larger than the characteristic scales predicted
by purely hydrodynamical models.  In particular, the absence of
bulk-Comptonization spectral signatures in blazars implies that Lorentz
factors $\gtrsim 10$ must be attained on scales not much smaller than
$\sim 10^{17}\,$cm.\cite{Sik05} This conclusion is supported by evidence
from radio VLBI observations for ongoing acceleration of distinct
outflow components on scales of $\sim 1-10\,$pc in many blazar
jets.\cite{Hom09} Observations of jets in sources like M87
indicate\cite{JBL99} that the bulk of the flow collimation occurs on
similar, comparatively large scales. Both of these findings are
consistent with the predictions of the ideal-MHD acceleration
model.\cite{VK04}
 
The superluminal radio components in blazar jets often appear to move on
helical paths. In the magnetic acceleration model, these observations
can be interpreted in terms of distinct outflow components that move
along helical field lines,\cite{CK92} with any given component
conceivably corresponding to an ejection episode along an isolated
magnetic flux bundle that threads the nuclear accretion disk. This
picture could be tested against alternative interpretations (involving
unstable fluid modes, source precession, etc.) by applying all the
available kinematic constraints from the VLBI observations. Valuable
additional constraints could be provided by the radiative properties of
the jets, including their radio and higher-energy
emission,\cite{Zen97}\cdash\cite{Mar08} linear and circular
polarization,\cite{LPG05}\cdash\cite{Gab08} and Faraday rotation
measure.\cite{BL09}

\section{Radiative Effects}

The nature of the radiative processes that give rise to the observed
emission from relativistic jets is clearly relevant to our ability to
properly interpret these sources. In fact, the lingering uncertainty on
this matter underlies some of the key remaining open questions in this
field, including the origin of the prompt and afterglow emission in
GRBs, the origin of the highest-energy $\gamma$-rays measured in GRBs
and in AGNs, and the origin of the extended X-ray emission from quasar
jets. As discussed below, these questions are closely linked to the
issue of the jet dynamics, both because the dynamical framework
influences which emission process dominates and because some of the
radiative effects have a direct dynamical impact on the flow. This
linkage can be manifested at the origin (e.g. when neutrino emission
from the accretion disk leads to an initially ``hot'' GRB outflow) or as
the jet makes its way out (e.g. the inverse-Compton interaction between
an AGN jet and the ambient radiation field), and it might even represent
a fundamental connection between the acceleration and emission
mechanisms (as when magnetic energy dissipation along the flow plays a
major role in both of these processes).

\subsection{Thermal acceleration effects}

If the enthalpy per rest-mass energy at the origin is $\gg 1$ then the
initial acceleration will be thermal, but the magnetic field will still
guide the flow if the Poynting/enthalpy flux ratio is $>1$ (or even if
this ratio is initially $< 1$ so long as the flow remains
sub-Alfv\'enic; Ref.~\refcite{VK03a}). Depending on the confining
pressure distribution and the adiabatic index of the gas, the enthalpy
flux in the thermal acceleration region could either go only into
kinetic energy or else be partially converted into Poynting flux
(thereby contributing to the jet acceleration further out).\cite{Kom09}
One effect of such added thermal acceleration would be to shift the
angular distribution of the jet kinetic power toward the axis, but a
clearer observational diagnostic of a ``fireball'' component would be
the presence of a photospheric thermal emission component in the
spectrum.\cite{DM02} Such a component is evidently missing in many GRBs
(including GRB 080916C; Ref.~\refcite{ZP09}) but appears to be present
in certain cases (e.g. GRB 090902B; Ref.~\refcite{Ryd10}). It may be
possible to explain this range of behaviors within the framework of the
basic dynamical model in terms of a variation in the enthalpy/rest-mass
and enthalpy/Poynting flux ratios among different sources.

\subsection{Interaction with the ambient radiation field}
\label{subsec:interact}

Focusing on the AGN case,\cite{SBR94} possible
sources of ambient photons include the nuclear disk emission, scattered
or reprocessed disk radiation, broad-line region emission, and the
microwave background. The most direct effect is inverse-Compton (i.c.)
scattering by electrons in the jet, which gives rise to the so-called
``external'' i.c. spectral component. For example, it has been
argued\cite{GMT09} that the apparent $\gamma$-ray spectral
separation within {\it Fermi}\/ blazars between BL Lac objects (which
have hard spectra) and flat-spectrum radio quasars (which have soft
high-energy spectra) could be attributed to the relative strength of
this component in these two types of sources. Furthermore, as already
noted in Sec.~\ref{subsec:AGN}, the constraint provided by the predicted
strength of this component places a lower bound on the spatial extent
of the jet acceleration region. This limit could be sharpened by
incorporating radiation drag and i.c. cooling explicitly into MHD
simulations of relativistic jets.

On the basis of taking into account also the possible conversion of
upscattered photons into $e^+e^-$ pairs through photon-photon
collisions, it was argued\cite{SP06}\cdash\cite{SP08} that a significant
fraction of the kinetic energy of AGN jets could potentially be
converted into high-energy emission. However, more elaborate numerical
simulations are needed to verify the viability of this {\it photon
breeding}\/ mechanism. In particular, two key assumptions of the model,
involving the dynamical response of the ambient medium and the nature of
its postulated magnetization, require further scrutiny.

\subsection{Uncertainties involving the emission mechanism}
\label{subsec:uncertain}

Concentrating on GRB sources, I address, in turn, questions about the
prompt and the afterglow emission components. It is noteworthy that the
origin of the prompt high-energy emission is still being debated: What
is the origin of the nonthermal component (synchrotron? jitter
radiation? synchrotron self-Compton?), and is there also a photospheric
thermal component present? Furthermore, the popular internal-shocks
model for the prompt emission has been challenged in several important
ways, including by indications from {\it Swift}\/ and {\it Fermi}\/
measurements that the emission region is much farther from the origin
($\gtrsim 10^{15}-10^{16}\,$cm; Refs.~\refcite{Lyu06}
and~\refcite{ZP09}) than what is expected in this scenario. One possible
alternative to this model is relativistic turbulence (with random
Lorentz factors $\sim 10$ in the comoving frame), as originally proposed
in the context of a force-free electrodynamics ($\rho_{\rm e} \mathbf{E}
+ \mathbf{j} \mathbf{\times} \mathbf{B} = 0$) outflow model.\cite{LB03}
GRB 080319B may be an example of a source where the data are consistent
with this picture but not with the internal-shocks
interpretation.\cite{KN09}

It is interesting to note in this connection that localized relativistic
emission regions that are distinct from the bulk jet outflow have also
been proposed to account for the TeV variability in
blazars.\cite{Ghi09b}\cdash\cite{GUB09} The latter proposals were
formulated in the context of the MHD dynamical model, which suggests
that a viable alternative to the internal-shocks model for GRBs need not
require replacing the MHD description by the force-free electrodynamic
(also referred to as the {\it magnetodynamic}) model. Magnetodynamics is
a limiting case of MHD when the matter inertia can be neglected, and it
allows one to derive the asymptotic parameter scalings of the full model
(Refs.~\refcite{Tch08} and~\refcite{Kom09}); however, one cannot
explicitly calculate the jet velocity or the magnetic-to-kinetic energy
conversion efficiency in this approximation.

Even if the MHD picture remains applicable, {\it ideal}\/ MHD may not
be. Highly relativistic Poynting flux-dominated outflows may be
inherently dissipative,\cite{Bla02} and this could lead to the direct
conversion of magnetic energy to radiation in both GRB and AGN
jets.\cite{Uso94}\cdash\cite{RL97} However, the specifics of the process
by which magnetic dissipation leads to particle acceleration and
nonthermal emission are not yet well understood and would need to be
considered in relation to alternative particle-acceleration mechanisms
that might operate in the jet, including processes that tap into the
cross-flow velocity shear (Refs.~\refcite{SO02} and~\refcite{LO07}).
Furthermore, this process might occur under certain constraints
(for example, the conservation of magnetic helicity) that could limit
the amount of dissipated energy and affect the internal magnetic field
structure.\cite{KC85}

Magnetic energy dissipation naturally creates a magnetic pressure
gradient along the flow, and it was proposed\cite{DS02} that even on its
own (i.e. without incorporating the acceleration that occurs in the
absence of dissipation) this effect could efficiently accelerate the jet
to relativistic speeds. However, so far this process has been modeled
only in an approximate manner, using strong simplifications: it needs to
be studied under more realistic assumptions with the help of {\it
resistive}\/ relativistic-MHD codes. It was also suggested that the
internal magnetic dissipation might be induced by current-driven
instabilities (in particular the $m=1$ kink),\cite{GS06} but the results
of recent MHD simulations indicate that the overall dynamics of
successfully launched relativistic jets is not strongly affected by such
instabilities\cite{MHN07}\cdash\cite{MB09} and that the associated
magnetic dissipation is correspondingly modest.\cite{MSO08} There is
observational evidence for strong dissipation in initially Poynting
flux-dominated outflows from pulsars, but how exactly this is achieved
(the ``$\sigma$ problem'' in pulsar-wind theory) and whether efficient
dissipation in relativistic jet sources requires a similar setup are
still open questions.

The origin of the afterglow emission also remains a puzzle. In
particular, the number of distinct outflow components observed
at any given phase of the afterglow evolution and the relative
contributions of the synchrotron and i.c. processes at any given
spectral regime are still open questions. Recent {\it Fermi}\/
observations were interpreted as indicating that the high-energy
($\gtrsim 10^2\,$MeV) photons are produced by the synchrotron process in
the forward shock that propagates into the ambient medium and that this
is also the source of the late ($\gtrsim 1\,$day) afterglow emission
but {\it not}\/ of the $< 10^2\,$MeV photons from the early
afterglow.\cite{KBD09}\cdash\cite{KBD10} In this picture, the inferred
magnetic field in the emission region is consistent with a
shock-compressed interstellar field, which mitigates the difficulty of
explaining the much stronger magnetic field deduced in the synchrotron
interpretation of the $< 10^2\,$MeV early afterglow.\cite{ML99} The
latter field can be naturally explained if the early afterglow emission
originates in the reverse shock that is driven into the (likely
magnetized) GRB ejecta, and this picture can in fact account for many
observed afterglow features.\cite{GDM07}\cdash\cite{UB07} An important
constraint on this scenario is provided by the fact that, if the jet
remains too strongly magnetized at the location of the early-afterglow
emission, the reverse shock may be too weak to account for the observed
radiation.\cite{MGA09} A further challenge is presented by the evidence
from {\it Swift}\/ observations that the optical and X-ray afterglow
light curves arise in two distinct emission
components.\cite{Ghi09a}\cdash\cite{Nar10} A complete model would need
to put all these pieces of the puzzle in place.

\section{The Accretion--Outflow (Disk--Jet) Connection}
\label{sec:disk_jet}

It has been known for some time now that the same scaling relation
between the radio ($L_{\rm R}$) and X-ray ($L_{\rm X}$) luminosities,
normalized by the black-hole mass $M$, applies in both AGNs and galactic
black-hole XRBs:\cite{MHD03}\cdash\cite{FKM04} \begin{equation}
\log{L_{\rm R}} \approx 0.60 \log{L_{\rm X}} + 0.78\log{M} + 7.33\; .
\nonumber \end{equation} The X-ray and radio luminosities can be
regarded as proxies for the accretion and outflow powers,
respectively. Interestingly, a positive correlation between the jet
power and the luminosity of the accretion disk is found also in
broad-line blazars when the $\gamma$-ray luminosity (inferred from {\it
Swift}\/ and {\it Fermi}\/ data) is used as a proxy for the jet
power.\cite{Ghi10} Black-hole XRBs exhibit this relation in the low/hard
state; the radio emission is quenched when the X-ray luminosity grows to
$\lesssim 10\%\; L_{\rm Edd}$ and the source enters the high/soft
state. A similar quenching of the radio emission is inferred in AGNs,
with sources such as Narrow-Line Seyfert 1 galaxies possibly corresponding
to the high/soft state.\cite{GHU06}

The low/hard state in galactic black-hole binaries has been interpreted
as an accretion phase during which steady-state jets carry away most of
the liberated power.\cite{LPK03} Transient jet outflows may occur at
higher accretion rates during the very-high (or steep-power-law) state,
of which powerful radio-jet sources may be the AGN analogs.\cite{Jes05}
It was proposed that jet dominance during the low/hard state might be
related to the formation of a large-scale poloidal field configuration,
although the details of this process and whether it is even the correct
explanation are still open questions. Compatibility with spectral
models, which indicate that the hard state requires a comparatively weak
magnetic field in the X-ray emission region and that the soft state is
actually consistent with a comparatively strong field,\cite{PV09} would
need to be part of the comprehensive picture that eventually emerges for
these sources.

Over the last several years, various studies have attempted to refine
and extend the empirical results for XRBs and to further clarify the
analogies between the accretion states of XRBs and AGNs (see
Refs.~\refcite{Fen10} and~\refcite{McH10} for reviews). Furthermore,
multi-frequency monitoring of the radio galaxy 3C 120 succeeded in
uncovering the same type of anti-correlated X-ray and radio variability
that is exhibited during radio-component ejections in XRBs.\cite{Cha09}
In considering these findings, it may be useful to recall that a strong
accretion--outflow connection is indicated in all cosmic jet sources
(including young stellar objects) but that its precise nature remains a
puzzle. Relativistic jet sources could potentially provide new insights
into this problem. For example, the extent to which the outflow is
influenced by the properties of the central object (rather than by those
of the accretion flow alone) is one important issue that could be
explicitly addressed in this case. In particular, one could
attempt to clarify the role of the black-hole spin in driving XRB
outflows by comparing the properties of neutron-star and black-hole
systems\cite{MF06} and then use the established analogies with AGNs to 
assess the applicability of the results to extragalactic jets.


\end{document}